\documentclass[aps, pra, showkeys, showpacs ]{revtex4}

\usepackage[small]{caption}
\usepackage{amsmath,amsfonts,amscd,amssymb,epsf, epsfig, mathrsfs,color}\usepackage{graphicx}

\usepackage{lscape,rotating}

\renewcommand{\d}{\operatorname{d}}

\newcommand{\pa}{\partial}

\newcommand{\vep}{\varepsilon}

\begin{document}

 \title{Casimir interaction between  spherical and planar plasma sheets }

\author{L. P. Teo}
 \email{LeePeng.Teo@nottingham.edu.my}
\affiliation{Department of Applied Mathematics, Faculty of Engineering, University of Nottingham Malaysia Campus, Jalan Broga, 43500, Semenyih, Selangor Darul Ehsan, Malaysia.}
\begin{abstract}We consider the interaction between a spherical plasma sheet and a planar plasma sheet due to the vacuum fluctuations of electromagnetic fields. We use the mode summation approach to derive the Casimir interaction energy and study its asymptotic behaviors. In the small separation regime, we confirm the proximity force approximation and calculate the first correction beyond the proximity force approximation. This study has potential application to model Casimir interaction between objects made of materials that can be modeled by plasma sheets such as graphene sheets.
\end{abstract}
\pacs{13.40.-f, 03.70.+k, 12.20.Ds}
\keywords{  Casimir interaction, sphere-plane configuration, plasma sheet, asymptotic behavior.}

\maketitle
\section{Introduction}

Due to the potential impact to nanotechnology, the Casimir interactions between objects of nontrivial geometries have been under active research in recent years. Thanks to works done by several groups of researchers \cite{1,2,3,4,5,6,7,8,9,10,11,12,13,14}, we have now a formalism to compute the exact functional representation (known as TGTG formula) for the Casimir interaction energy between two objects. Despite the seemingly different     approaches taken, all the methods can be regarded as multiple scattering approach, which can also be understood from the point of view of mode summation approach \cite{43,44,45}. The basic ingredients in the TGTG formula are the scattering matrices of the two objects and the transition matrices that relate the coordinate system of one object to the other. In the case that the objects  have certain symmetries that allow    separable coordinate system to be employed, one can calculate these matrices explicitly. This has made possible the exact analytic and numerical analysis of the Casimir interaction between a sphere and a plate \cite{34, 33, 32,29, 28, 35, 27, 26, 25, 15, 16, 19}, between two spheres \cite{31,23,22}, between a cylinder and a plate \cite{9,39,40}, between two cylinders \cite{42,38, 21,20}, between a sphere and a cylinder \cite{17, 36}, as well as other geometries \cite{30, 37, 41}.

As is well known, the strength of the Casimir interaction does not only depend on the   geometries of the objects, it is also very sensitive to the boundary conditions imposed on the objects. For the past few years, a lot of works have been done in the analysis of the quantum effect on objects with perfect boundary conditions such as Dirichlet, Neumann, perfectly conducting, infinitely permeable, etc. There are also a number of works which consider real materials such as  metals modeled by plasma or Drude models \cite{33,32,27,26,25,19,15,23,20,36}. In this work, we consider the Casimir interaction between a spherical plasma sheet and a planar plasma sheet. Plasma sheet model was considered  in   \cite{40, 46,47, 48, 49, 24, 18} to model graphene sheet, describing the $\pi$ electrons in C$_{60}$ molecule. This model has its own appeal in describing a thin shell of materials that have the same attributes.

In \cite{40}, the Casimir interaction between a cylindrical  plasma sheet and a planar plasma sheet has been considered. Our work can be considered as a generalization of \cite{40} where we consider a spherical plasma sheet instead of a cylindrical plasma sheet. One of the main objectives of the current work is to derive the TGTG formula for the Casimir interaction energy. As in \cite{40}, we are also going to study the asymptotic behaviors of the Casimir interaction in the small separation regimes. We would expect that  the leading term of the Casimir interaction coincides with the proximity force approximation (PFA), which   we are going to confirm. Another major contribution would be the exact analytic computation of the next-to-leading order term which determines the deviation from PFA.

\section{The Casimir interaction energy}

In this section, we derive the TGTG formula for the  Casimir interaction energy between a spherical plasma sheet and a planar plasma sheet. We follow our approach in \cite{45}.

Assume that the spherical plasma sheet is a spherical surface described by $r=R$ in spherical coordinates $(r,\theta,\phi)$, and the planar plasma sheet is located at $z=L$ with dimension $H\times H$. It is assumed that $R<L\ll H$. The center of the spherical shell is the origin $O$, and the center of the coordinate system about the plane $z=L$ is $O' =(0,0,L)$.

The  electromagnetic field is governed by the Maxwell's equations:
\begin{equation}\label{eq2_19_1}\begin{split}
&\nabla\cdot\mathbf{E}=\frac{\rho_f}{\vep_0},\hspace{1cm}\nabla\times\mathbf{E}+\frac{ \pa\mathbf{B} }{\pa t}=\mathbf{0},\\
&\nabla\cdot\mathbf{B}=0,\hspace{1cm} \nabla\times\mathbf{B}-\frac{1}{c^2}\frac{\pa\mathbf{E}}{\pa t}=\mu_0\mathbf{J}_f.
\end{split}\end{equation} The free charge density $\rho_f$ and free current density $\mathbf{J}_f$ are functions having support on the plasma sheets (boundaries). Let $\mathbf{A}$ be a vector potential that satisfies the gauge condition $\nabla\cdot\mathbf{A}=0$ and such that
\begin{equation*}
\mathbf{E}=-\frac{\pa\mathbf{A}}{\pa t},\hspace{1cm}\mathbf{B}=\nabla\times\mathbf{A}.
\end{equation*}$\mathbf{A}(\mathbf{x},t)$ can be written as a superposition of normal modes $\mathbf{A}(\mathbf{x},\omega)e^{-i\omega t}$:
$$\mathbf{A}(\mathbf{x},t)=\int_{-\infty}^{\infty}d\omega \,\mathbf{A}(\mathbf{x},\omega)e^{-i\omega t}.$$
Maxwell's equations \eqref{eq2_19_1} imply that outside the boundaries,
\begin{equation}\label{eq2_18_1}
\nabla\times\nabla \times \mathbf{A}(\mathbf{x},\omega)=k^2\mathbf{A}(\mathbf{x},\omega),
\end{equation}where
$$k=\frac{\omega}{c}.$$
 The boundary conditions are given by \cite{46}:
\begin{equation}\label{eq2_18_2}\begin{split}
&\mathbf{E}_{\parallel}\Bigr|_{S_+}-\mathbf{E}_{\parallel}\Bigr|_{S_-}
=\mathbf{0},\\
&\mathbf{B}_{n}\Bigr|_{S_+}-\mathbf{B}_{n}\Bigr|_{S_-}=0,\\
&\mathbf{E}_{n}\Bigr|_{S_+}-\mathbf{E}_{n}\Bigr|_{S_-}=2\Omega\frac{c^2}{\omega^2}\nabla_{\parallel}\cdot\mathbf{E}_{\parallel}\Bigr|_{ S},\\
&\mathbf{B}_{\parallel}\Bigr|_{S_+}-\mathbf{B}_{\parallel}\Bigr|_{S_-}
=-2i\Omega\frac{1}{\omega}\mathbf{n}\times\mathbf{E}_{\parallel}\Bigr|_{S},
\end{split}\end{equation}where $S$ is the boundary, $S_+$ and $S_-$ are respectively the outside and inside of the boundary, $\mathbf{n}$ is a unit vector normal to the boundary, and $\Omega$ is a constant characterizing the plasma, having dimension inverse of length.

The solutions of the equation \eqref{eq2_18_1} can be   divided  into transverse electric (TE) waves $\mathbf{A}^{\text{TE}}_{\alpha}$ and transverse magnetic (TM)
waves $\mathbf{A}^{\text{TM}}_{\alpha}$ parametrized by some parameter $\alpha$ and satisfy
\begin{equation}\label{eq2_18_3}
\frac{1}{k}\nabla\times \mathbf{A}^{\text{TE}}_{\alpha}=\mathbf{A}^{\text{TM}}_{\alpha},\hspace{1cm}\frac{1}{k}\nabla\times \mathbf{A}^{\text{TM}}_{\alpha}=
\mathbf{A}^{\text{TE}}_{\alpha}.
\end{equation}Moreover, the waves can be divided into regular waves $\mathbf{A}^{\text{TE,reg}}_{\alpha}$, $\mathbf{A}^{\text{TM,reg}}_{\alpha}$ that are
regular at the origin of the coordinate system  and outgoing waves $\mathbf{A}^{\text{TE,out}}_{\alpha}$, $\mathbf{A}^{\text{TM,out}}_{\alpha}$ that decrease
to zero rapidly when $\mathbf{x}\rightarrow  \infty$ and $k$ is replaced by $ik$.

In rectangular coordinates, the waves are parametrized by $\alpha=\mathbf{k}_{\perp}=(k_x,k_y)\in\mathbb{R}^2$, with  \begin{equation*}\begin{split}
\mathbf{A}_{\mathbf{k}_{\perp}}^{\text{TE}, \substack{\text{reg}\\\text{out}}}(\mathbf{x},\omega)
=& \frac{1}{k_{\perp}}e^{ik_xx+ik_yy\mp i\sqrt{k^2-k_{\perp}^2}z}
\left(ik_y\mathbf{e}_x-ik_x\mathbf{e}_y\right),\\
\mathbf{A}_{\mathbf{k}_{\perp}}^{\text{TM}, \substack{\text{reg}\\\text{out}}}(\mathbf{x},\omega)
=& \frac{1}{kk_{\perp}}e^{ik_xx+ik_yy\mp i\sqrt{k^2-k_{\perp}^2}z}
\left(\pm k_x\sqrt{k^2-k_{\perp}^2}\mathbf{e}_x \pm k_y\sqrt{k^2-k_{\perp}^2}\mathbf{e}_y+k_{\perp}^2\mathbf{e}_z\right).
\end{split}\end{equation*}
Here $k_{\perp}=\sqrt{k_x^2+k_y^2}$.

In spherical coordinates, the waves are parametrized by $\alpha=(l,m)$, where $l=1,2,3,\ldots$ and $-l\leq m\leq l$, with
\begin{equation*}\begin{split}
\mathbf{A}_{lm}^{\text{TE},*}(\mathbf{x},\omega)
=&\frac{\mathcal{C}_{l}^*}{\sqrt{l(l+1)}}f_l^*(kr)\left(\frac{im}{\sin\theta}Y_{lm}(\theta,\phi)\mathbf{e}_{\theta}-\frac{\pa Y_{lm}(\theta,\phi)}{\pa\theta}
\mathbf{e}_{\phi}\right),\\
\mathbf{A}_{lm}^{\text{TM},*}(\mathbf{x},\omega)
=&\mathcal{C}_{l}^*\left(\frac{\sqrt{l(l+1)}}{kr}f_l^*(kr)Y_{lm}(\theta,\phi)\mathbf{e}_r+\frac{1}{\sqrt{l(l+1)}}\frac{1}{kr}\frac{d}{dr}\left(rf_l^*(kr)\right)
\left[\frac{\pa Y_{lm}(\theta,\phi)}{\pa\theta}\mathbf{e}_{\theta}+\frac{im}{\sin\theta}Y_{lm}(\theta,\phi)\mathbf{e}_{\phi}\right]\right).
\end{split}\end{equation*}Here $*=$ reg or out, with $f_l^{\text{reg}}(z)=j_l(z)$ and $f_l^{\text{out}}(z)=h_l^{(1)}(z)$, $Y_{lm}(\theta,\phi)$ are the spherical harmonics. The constants $\mathcal{C}_{l}^{\text{reg}}$ and $\mathcal{C}_{l}^{\text{out}}$ are chosen so that
\begin{equation*}
\mathcal{C}_{l}^{\text{reg}} j_l(i\zeta) =\sqrt{\frac{\pi}{2\zeta}}I_{l+\frac{1}{2}}(\zeta),\quad \mathcal{C}_{l}^{\text{out}} h_l^{(1)}(i\zeta) =\sqrt{\frac{\pi}{2\zeta}}K_{l+\frac{1}{2}}(\zeta).
\end{equation*}

Now we can derive the dispersion relation for the energy eigenmodes $\omega$ of the system. Inside the sphere ($r<R$),   express $\mathbf{A}(\mathbf{x},t) $ in the spherical coordinate system centered at $O$:
\begin{equation*}
\mathbf{A}(\mathbf{x},t) =\int_{-\infty}^{\infty} d\omega \sum_{l=1}^{\infty}\sum_{m=-l}^l\left(A_1^{lm} \mathbf{A}^{\text{TE,reg}}_{lm}(\mathbf{x},\omega)+C_1^{lm}
\mathbf{A}^{\text{TM,reg}}_{lm}(\mathbf{x},\omega)\right)e^{-i\omega t}.
\end{equation*}
Outside the plane ($z>L$),
  express $\mathbf{A} $ in the rectangular coordinate system centered at $O'$:
\begin{equation*}
\mathbf{A}(\mathbf{x}',t) =H^2\int_{-\infty}^{\infty} d\omega \int_{-\infty}^{\infty}\frac{dk_x}{2\pi}\int_{-\infty}^{\infty}\frac{dk_y}{2\pi} \left(B_2^{\mathbf{k}_{\perp}} \mathbf{A}^{\text{TE,out}}_{\mathbf{k}_{\perp}}(\mathbf{x}',\omega)+D_2^{\mathbf{k}_{\perp}}
\mathbf{A}^{\text{TM,out}}_{\mathbf{k}_{\perp}}(\mathbf{x}',\omega)\right)e^{-i\omega t}.
\end{equation*}
Here $\mathbf{x}'=\mathbf{x}-\mathbf{L}$, $\mathbf{L}=L\mathbf{e}_z$.

In the region between the sphere and the plane, $\mathbf{A}$ can be represented in two ways: one is in terms of the spherical coordinate system centered at $O$:
\begin{equation*}
\mathbf{A}(\mathbf{x},t) =\int_{-\infty}^{\infty} d\omega \sum_{l=1}^{\infty}\sum_{m=-l}^l\left(a_1^{lm} \mathbf{A}^{\text{TE,reg}}_{lm}(\mathbf{x},\omega)+
b_1^{lm} \mathbf{A}^{\text{TE,out}}_{lm}(\mathbf{x},\omega)+c_1^{lm} \mathbf{A}^{\text{TM,reg}}_{lm}(\mathbf{x},\omega)+d_1^{lm}
\mathbf{A}^{\text{TM,out}}_{lm}(\mathbf{x},\omega)\right)e^{-i\omega t};
\end{equation*}and one is in terms of the rectangular coordinate system centered at $O'$:
\begin{equation*}\begin{split}
\mathbf{A}(\mathbf{x}',t) =&H^2\int_{-\infty}^{\infty} d\omega \int_{-\infty}^{\infty}\frac{dk_x}{2\pi}\int_{-\infty}^{\infty}\frac{dk_y}{2\pi}\\&\hspace{2cm}\times\left(a_2^{\mathbf{k}_{\perp} }\mathbf{A}^{\text{TE,reg}}_{\mathbf{k}_{\perp} }(\mathbf{x}',\omega)+
b_2^{\mathbf{k}_{\perp} } \mathbf{A}^{\text{TE,out}}_{\mathbf{k}_{\perp}}(\mathbf{x}',\omega)+c_2^{\mathbf{k}_{\perp} } \mathbf{A}^{\text{TM,reg}}_{\mathbf{k}_{\perp} }(\mathbf{x}',\omega)+d_2^{\mathbf{k}_{\perp} }
\mathbf{A}^{\text{TM,out}}_{\mathbf{k}_{\perp} }(\mathbf{x}',\omega)\right)e^{-i\omega t}.\end{split}
\end{equation*}These two representations are related by translation matrices $\mathbb{V}$ and $\mathbb{W}$:
\begin{equation*}\begin{split}
\begin{pmatrix}\mathbf{A}^{\text{TE,reg}}_{\mathbf{k}_{\perp}}(\mathbf{x}',\omega)\\\mathbf{A}^{\text{TM,reg}}_{\mathbf{k}_{\perp}}(\mathbf{x}',\omega)\end{pmatrix}
=&\sum_{l=1}^{\infty}\sum_{m=-l}^l \begin{pmatrix}V_{lm, \mathbf{k}_{\perp}}^{\text{TE,TE}}  & V_{ lm, \mathbf{k}_{\perp}}^{\text{TM,TE}} \\
 V_{ lm, \mathbf{k}_{\perp}}^{\text{TE,TM}}  & V_{lm, \mathbf{k}_{\perp}}^{\text{TM,TM}} \end{pmatrix}
 \begin{pmatrix}\mathbf{A}^{\text{TE,reg}}_{lm}(\mathbf{x},\omega)
\\\mathbf{A}^{\text{TM,reg}}_{lm}(\mathbf{x},\omega)\end{pmatrix}\\
\begin{pmatrix}\mathbf{A}^{\text{TE,out}}_{lm}(\mathbf{x},\omega)\\\mathbf{A}^{\text{TM,out}}_{lm}(\mathbf{x},\omega)\end{pmatrix}
=&H^2\int_{-\infty}^{\infty}\frac{dk_x}{2\pi}\int_{-\infty}^{\infty}\frac{dk_y}{2\pi} \begin{pmatrix}W_{\mathbf{k}_{\perp}, lm}^{\text{TE,TE}}  & W_{\mathbf{k}_{\perp}, lm}^{\text{TM,TE}} \\
W_{\mathbf{k}_{\perp}, lm}^{\text{TE,TM}}  & W_{\mathbf{k}_{\perp}, lm}^{\text{TM,TM}}  \end{pmatrix}
 \begin{pmatrix}\mathbf{A}^{\text{TE,out}}_{\mathbf{k}_{\perp}}(\mathbf{x}',\omega)
\\\mathbf{A}^{\text{TM,out}}_{\mathbf{k}_{\perp}}(\mathbf{x}',\omega)\end{pmatrix}.
\end{split}\end{equation*}
Hence,
\begin{equation*}\begin{split}
\begin{pmatrix}
a_1^{lm}\\c_1^{lm}
\end{pmatrix}=&H^2\int_{-\infty}^{\infty}\frac{dk_x}{2\pi}\int_{-\infty}^{\infty}\frac{dk_y}{2\pi}\begin{pmatrix}V_{lm, \mathbf{k}_{\perp}}^{\text{TE,TE}}  & V_{ lm, \mathbf{k}_{\perp}}^{\text{TE,TM}}  \\
 V_{ lm, \mathbf{k}_{\perp}}^{\text{TM,TE}}  & V_{lm, \mathbf{k}_{\perp}}^{\text{TM,TM}}  \end{pmatrix}\begin{pmatrix}
a_2^{\mathbf{k}_{\perp}}\\c_2^{\mathbf{k}_{\perp}}
\end{pmatrix},\\
\begin{pmatrix}
b_2^{\mathbf{k}_{\perp}}\\d_2^{\mathbf{k}_{\perp}}
\end{pmatrix}=&\sum_{l=1}^{\infty}\sum_{m=-l}^l\begin{pmatrix}W_{\mathbf{k}_{\perp}, lm}^{\text{TE,TE}}  & W_{\mathbf{k}_{\perp}, lm}^{\text{TE,TM}}  \\
W_{\mathbf{k}_{\perp}, lm}^{\text{TM,TE}}  & W_{\mathbf{k}_{\perp}, lm}^{\text{TM,TM}} \end{pmatrix}\begin{pmatrix}
b_1^{lm}\\d_1^{lm}
\end{pmatrix}.\end{split}
\end{equation*}
These translation matrices have been derived in \cite{8,45}. Their components are given by
\begin{equation*}
\begin{split}
 V_{lm,\mathbf{k}_{\perp}}^{\text{TE,TE}} =V_{lm,\mathbf{k}_{\perp}}^{\text{TM,TM}}
 =&-\frac{4\pi i }{\sqrt{l(l+1)}}  \frac{\pa Y_{l,-m}(\theta_k,\phi_k)}{\pa\theta_k}e^{i\sqrt{k^2-k_{\perp}^2}L},\\
 V_{lm,\mathbf{k}_{\perp}}^{\text{TE,TM}} =V_{lm,\mathbf{k}_{\perp}}^{\text{TM,TE}} =&\frac{4\pi i  }{\sqrt{l(l+1)}} \frac{m}{\sin\theta_k} Y_{l,-m}(\theta_k,\phi_k) e^{i\sqrt{k^2-k_{\perp}^2}L},
\end{split}
\end{equation*}
\begin{equation*}
\begin{split}
W_{\mathbf{k}_{\perp},lm}^{\text{TE,TE}} =&\frac{i }{H^2\sqrt{l(l+1)}} \frac{\pi^2}{k\sqrt{k^2-k_{\perp}^2}}
\frac{\pa Y_{lm}(\theta_k,\phi_k)}{\pa\theta_k}e^{i\sqrt{k^2-k_{\perp}^2}L},\\
W_{\mathbf{k}_{\perp},lm}^{\text{TM,TE}} =&\frac{i }{H^2\sqrt{l(l+1)}} \frac{\pi^2}{k\sqrt{k^2-k_{\perp}^2}}
\frac{m}{\sin\theta_k} Y_{lm}(\theta_k,\phi_k) e^{i\sqrt{k^2-k_{\perp}^2}L}.
\end{split}
\end{equation*}Here $\theta_k$ and $\phi_k$ are such that $k_{\perp}=k\sin\theta_k$, $k_x=k_{\perp}\cos\phi_k$ and $k_y=k_{\perp}\sin\phi_k$.

Let $\Omega_s$ be the parameter characterizing the spherical plasma sheet. Matching the boundary conditions \eqref{eq2_18_2} on the sphere gives
\begin{equation*}
\begin{split}
&a_1^{lm}\mathcal{C}_{l }^{\text{reg}}j_l(kR)+b_1^{lm}\mathcal{C}_{l }^{\text{out}}h_l^{(1)}(kR)=A_1^{lm}\mathcal{C}_{l }^{\text{reg}}j_l(kR),\\
&a_1^{lm}\mathcal{C}_{l }^{\text{reg}} \Bigl(j_l(kR)+kRj_l'(kR)\Bigr)+b_1^{lm}\mathcal{C}_{l }^{\text{out}}
\left(h_l^{(1)}(kR)+kRh_l^{(1)\prime}(kR)\right)- A_1^{lm} \mathcal{C}_{l }^{\text{reg}}\Bigl(j_l(kR)+kRj_l'(kR)\Bigr)\\&\hspace{4cm}=2\Omega_s RA_1^{lm}
\mathcal{C}_{l }^{\text{reg}}j_l(kR),\\
&c_1^{lm}\mathcal{C}_{l }^{\text{reg}}\Bigl(j_l(kR)+kRj_l'(kR)\Bigr)+d_1^{lm}\mathcal{C}_{l }^{\text{out}}
\left(h_l^{(1)}(kR)+kRh_l^{(1)\prime}(kR)\right)=C_1^{lm}\mathcal{C}_{l }^{\text{reg}}\Bigl(j_l(kR)+kRj_l'(kR)\Bigr),\\
&c_1^{lm}\mathcal{C}_{l }^{\text{reg}}j_l(kR)+d_1^{lm}\mathcal{C}_{l }^{\text{out}}h_l^{(1)}(kR)-C_1^{lm}\mathcal{C}_{l }^{\text{reg}}j_l(kR)=
-\frac{2\Omega_s c^2}{\omega^2R}C_1^{lm}\mathcal{C}_{l }^{\text{reg}}\Bigl(j_l(kR)+kRj_l'(kR)\Bigr).
\end{split}\end{equation*}Eliminating $A^{lm}$ and $C^{lm}$, we obtain a relation of the form
\begin{align*}
\begin{pmatrix} b_1^{lm}\\d_1^{lm}\end{pmatrix}=-\mathbb{T}_{lm}\begin{pmatrix} a_1^{lm}\\c_1^{lm}\end{pmatrix},
\end{align*}where $\mathbb{T}_{lm}$ is a diagonal matrix:
\begin{equation*}
\mathbb{T}_{lm}=\begin{pmatrix} T_{lm}^{\text{TE}}&0\\0& T_{lm}^{\text{TM}}\end{pmatrix}
\end{equation*}with
\begin{equation}\label{eq2_18_5}\begin{split}
T_{lm}^{\text{TE}}(i\xi)=& \frac{2\Omega_s R I_{l+\frac{1}{2}}(\kappa R)^2}{1+2\Omega_s RI_{l+\frac{1}{2}}(\kappa R)K_{l+\frac{1}{2}}(\kappa R)},\\
T_{lm}^{\text{TM}}(i\xi)=& -\frac{ 2\Omega_s  \left(\frac{1}{2}I_{l+\frac{1}{2}}(\kappa R)+
\kappa RI_{l+\frac{1}{2}}'(\kappa R)\right)^2}{\kappa^2R- 2\Omega_s \left(\frac{1}{2}I_{l+\frac{1}{2}}(\kappa R)+
\kappa RI_{l+\frac{1}{2}}'(\kappa R)\right)\left(\frac{1}{2}K_{l+\frac{1}{2}}(\kappa R)+
\kappa RK_{l+\frac{1}{2}}'(\kappa R)\right)}.
\end{split}
\end{equation}Here we have replaced $k$ by $i\kappa$ and $\omega$ by $i\xi$.

Denote by $\Omega_p$ be the parameter characterizing the planar plasma sheet. Matching the boundary conditions \eqref{eq2_18_2} on the plane gives
\begin{equation*}
\begin{split}
&a_2^{\mathbf{k}_{\perp}}+ b_2^{\mathbf{k}_{\perp}}=B_2^{\mathbf{k}_{\perp}},\\
&\sqrt{k^2-k_{\perp}^2}\left(a_2^{\mathbf{k}_{\perp}}- b_2^{\mathbf{k}_{\perp}}+B_2^{\mathbf{k}_{\perp}}\right)=-2i\Omega_p B_2^{\mathbf{k}_{\perp}},\\
&c_2^{\mathbf{k}_{\perp}}-d_2^{\mathbf{k}_{\perp}}=-D_2^{\mathbf{k}_{\perp}},\\
&c_2^{\mathbf{k}_{\perp}}+d_2^{\mathbf{k}_{\perp}}-D_2^{\mathbf{k}_{\perp}}=\frac{2i\Omega_p c^2}{\omega^2}\sqrt{k^2-k_{\perp}^2}D_2^{\mathbf{k}_{\perp}}.
\end{split}
\end{equation*}From here, we find that   \begin{align*}
\begin{pmatrix} a_2^{\mathbf{k}_{\perp}}\\c_2^{\mathbf{k}_{\perp}}\end{pmatrix}=-\widetilde{\mathbb{T}}_{\mathbf{k}_{\perp}}\begin{pmatrix} b_2^{\mathbf{k}_{\perp}}\\d_2^{\mathbf{k}_{\perp}}\end{pmatrix},
\end{align*} where $\widetilde{\mathbb{T}}_{\mathbf{k}_{\perp}}$ is a diagonal matrix with elements
\begin{equation}\label{eq2_19_4}
\begin{split}
\widetilde{T}_{\mathbf{k}_{\perp}}^{\text{TE}}(i\xi)=&\frac{\Omega_p}{\Omega_p+\sqrt{\kappa^2+k_{\perp}^2}},\\
\widetilde{T}_{\mathbf{k}_{\perp}}^{\text{TM}}(i\xi)=&-\frac{\Omega_p \sqrt{\kappa^2+k_{\perp}^2}}{\Omega_p \sqrt{\kappa^2+k_{\perp}^2}+\kappa^2}.
\end{split}
\end{equation}

The eigenmodes $\omega$ are those modes where the boundary conditions give rise to nontrivial solutions of $(A_1^{lm}, C_1^{lm}, B_2^{\mathbf{k}_{\perp}}, D_2^{\mathbf{k}_{\perp}}, a_1^{lm}, b_1^{lm}, c_1^{lm}, d_1^{lm}, a_2^{\mathbf{k}_{\perp}}, b_2^{\mathbf{k}_{\perp}}, c_2^{\mathbf{k}_{\perp}}, d_2^{k_{\perp}})$. Now
\begin{align*}
\begin{pmatrix} b_1^{lm}\\d_1^{lm}\end{pmatrix}=&-\mathbb{T}_{lm}\begin{pmatrix} a_1^{lm}\\c_1^{lm}\end{pmatrix}\\
=&-\mathbb{T}_{lm} H^2\int_{-\infty}^{\infty}\frac{dk_x}{2\pi}\int_{-\infty}^{\infty}\frac{dk_y}{2\pi}\begin{pmatrix}V_{lm, \mathbf{k}_{\perp}}^{\text{TE,TE}}  & V_{ lm, \mathbf{k}_{\perp}}^{\text{TE,TM}}  \\
 V_{ lm, \mathbf{k}_{\perp}}^{\text{TM,TE}}  & V_{lm, \mathbf{k}_{\perp}}^{\text{TM,TM}}  \end{pmatrix}\begin{pmatrix}
a_2^{\mathbf{k}_{\perp}}\\c_2^{\mathbf{k}_{\perp}}\end{pmatrix}\\
=&\mathbb{T}_{lm} H^2\int_{-\infty}^{\infty}\frac{dk_x}{2\pi}\int_{-\infty}^{\infty}\frac{dk_y}{2\pi}\begin{pmatrix}V_{lm, \mathbf{k}_{\perp}}^{\text{TE,TE}}  & V_{ lm, \mathbf{k}_{\perp}}^{\text{TE,TM}}  \\
 V_{ lm, \mathbf{k}_{\perp}}^{\text{TM,TE}}  & V_{lm, \mathbf{k}_{\perp}}^{\text{TM,TM}}  \end{pmatrix}\widetilde{\mathbb{T}}_{\mathbf{k}_{\perp}}\begin{pmatrix} b_2^{\mathbf{k}_{\perp}}\\d_2^{\mathbf{k}_{\perp}}
\end{pmatrix}\\
 =&\mathbb{T}_{lm} H^2\int_{-\infty}^{\infty}\frac{dk_x}{2\pi}\int_{-\infty}^{\infty}\frac{dk_y}{2\pi}\begin{pmatrix}V_{lm, \mathbf{k}_{\perp}}^{\text{TE,TE}}  & V_{ lm, \mathbf{k}_{\perp}}^{\text{TE,TM}}  \\
 V_{ lm, \mathbf{k}_{\perp}}^{\text{TM,TE}}  & V_{lm, \mathbf{k}_{\perp}}^{\text{TM,TM}}  \end{pmatrix}\widetilde{\mathbb{T}}_{\mathbf{k}_{\perp}}\sum_{l'=1}^{\infty}\sum_{m'=-l'}^{l'}\begin{pmatrix}W_{\mathbf{k}_{\perp}, l'm'}^{\text{TE,TE}}  & W_{\mathbf{k}_{\perp}, l'm'}^{\text{TE,TM}}  \\
W_{\mathbf{k}_{\perp}, l'm'}^{\text{TM,TE}}  & W_{\mathbf{k}_{\perp}, l'm'}^{\text{TM,TM}} \end{pmatrix}\begin{pmatrix}
b_1^{l'm'}\\d_1^{l'm'}
\end{pmatrix}.
\end{align*}
This shows that the matrix $\mathbb{B}$ with $(lm)$ component given by
\begin{align*}
\begin{pmatrix} b_1^{lm}\\d_1^{lm}\end{pmatrix}
\end{align*} satisfies the relation
\begin{align*}
\left(\mathbb{I}-\mathbb{M}\right)\mathbb{B}=\mathbb{O},
\end{align*}where the $(lm, l'm')$-element of $\mathbb{M}$ is given by
\begin{align*}
M_{lm, l'm'}=\mathbb{T}_{lm} H^2\int_{-\infty}^{\infty}\frac{dk_x}{2\pi}\int_{-\infty}^{\infty}\frac{dk_y}{2\pi}\begin{pmatrix}V_{lm, \mathbf{k}_{\perp}}^{\text{TE,TE}}  & V_{ lm, \mathbf{k}_{\perp}}^{\text{TE,TM}}  \\
 V_{ lm, \mathbf{k}_{\perp}}^{\text{TM,TE}}  & V_{lm, \mathbf{k}_{\perp}}^{\text{TM,TM}}  \end{pmatrix}\widetilde{\mathbb{T}}_{\mathbf{k}_{\perp}} \begin{pmatrix}W_{\mathbf{k}_{\perp}, l'm'}^{\text{TE,TE}}  & W_{\mathbf{k}_{\perp}, l'm'}^{\text{TE,TM}}  \\
W_{\mathbf{k}_{\perp}, l'm'}^{\text{TM,TE}}  & W_{\mathbf{k}_{\perp}, l'm'}^{\text{TM,TM}} \end{pmatrix}.
\end{align*}The condition for nontrivial solution of $\mathbb{B}$ is thus given by
\begin{align*}
\det\left(\mathbb{I}-\mathbb{M}\right)=0.
\end{align*} Hence, the Casimir interaction energy between the spherical plasma sheet and the planar plasma sheet is
\begin{align}\label{eq2_19_2}
E_{\text{Cas}}
=\frac{\hbar}{2\pi}\int_0^{\infty} d\xi \text{Tr}\,\ln  \left(\mathbb{I}-\mathbb{M}(i\xi)\right)=\frac{\hbar c}{2\pi}\int_0^{\infty} d\kappa \text{Tr}\,\ln  \left(\mathbb{I}-\mathbb{M} \right).
\end{align}Set $k_x=k_{\perp}\cos\theta_k$, $k_y=k_{\perp}\sin\theta_k$, integrate over $\theta_k$ and make a change of variables $k_{\perp}=\kappa\sinh\theta$, we find that
 \begin{equation}\label{eq3_3_1}\begin{split}
\mathbb{M}_{lm,l'm'}(i\xi)=&\delta_{m,m'} \frac{(-1)^{m}\pi}{2}\sqrt{\frac{(2l+1)(2l'+1)}{l(l+1)l'(l'+1)}\frac{(l-m)!(l'-m)!}{(l+m)!(l'+m)!}}   \mathbb{T}_{lm}
 \int_{0}^{\infty}d\theta \sinh\theta e^{-2\kappa L\cosh\theta}\\&
\times
 \left(\begin{aligned} \sinh\theta P_l^{m\prime}(\cosh\theta)\hspace{0.5cm} &-\frac{m}{\sinh\theta}P_l^m(\cosh\theta)\\ -\frac{m}{\sinh\theta}P_l^m(\cosh\theta)
 \hspace{0.4cm} & \quad\sinh\theta P_l^{m\prime}(\cosh\theta) \end{aligned}\right)\left(\begin{aligned} \frac{\Omega_p}{\Omega_p+\kappa\cosh\theta} & ~\hspace{1cm} 0\hspace{0.5cm}\\
 \hspace{0.5cm} 0\hspace{1cm} & -\frac{\Omega_p   \cosh\theta}{\Omega_p \cosh\theta+\kappa}\end{aligned}\right)
 \\&\times\left(\begin{aligned} \sinh\theta P_{l'}^{m'\prime}(\cosh\theta)\hspace{0.5cm} & \frac{m'}{\sinh\theta}P_{l'}^{m'}(\cosh\theta)\\  \frac{m'}{\sinh\theta}P_{l'}^{m'}(\cosh\theta)
 \hspace{0.4cm} & \quad\sinh\theta P_{l'}^{m'\prime}(\cosh\theta) \end{aligned}\right).
\end{split}\end{equation}Here
$P_l^m(z)$ is an associated Legendre function
 and $P_l^{m\prime}(z)$ is its derivative, whereas $\mathbb{T}_{lm}$ is given by \eqref{eq2_18_5}.
 
 Notice that   this approach has been formalized mathematically in \cite{45}. The self energy contributions from the sphere and the plane have automatically dropped out and \eqref{eq2_19_2} is the interaction energy between the sphere and the plane.

In the limit $\Omega_s\rightarrow\infty$ and $\Omega_p\rightarrow\infty$, we find from \eqref{eq2_18_5} and \eqref{eq3_3_1} that
\begin{equation}\label{eq2_18_5_2}\begin{split}
T_{lm}^{\text{TE}}(i\xi)=& \frac{ I_{l+\frac{1}{2}}(\kappa R)}{K_{l+\frac{1}{2}}(\kappa R)},\\
T_{lm}^{\text{TM}}(i\xi)=& \frac{\frac{1}{2}I_{l+\frac{1}{2}}(\kappa R)+
\kappa RI_{l+\frac{1}{2}}'(\kappa R)}{\frac{1}{2}K_{l+\frac{1}{2}}(\kappa R)+
\kappa RK_{l+\frac{1}{2}}'(\kappa R)},
\end{split}
\end{equation}
\begin{equation*}\begin{split}
\mathbb{M}_{lm,l'm'}(i\xi)=&\delta_{m,m'} \frac{(-1)^{m}\pi}{2}\sqrt{\frac{(2l+1)(2l'+1)}{l(l+1)l'(l'+1)}\frac{(l-m)!(l'-m)!}{(l+m)!(l'+m)!}}   \mathbb{T}_{lm}
 \int_{0}^{\infty}d\theta \sinh\theta e^{-2\kappa L\cosh\theta}\\&
\times \begin{pmatrix} 1 & 0\\0& -1\end{pmatrix}
 \left(\begin{aligned} \sinh\theta P_l^{m\prime}(\cosh\theta)\hspace{0.5cm} &\frac{m}{\sinh\theta}P_l^m(\cosh\theta)\\ \frac{m}{\sinh\theta}P_l^m(\cosh\theta)
 \hspace{0.4cm} & \quad\sinh\theta P_l^{m\prime}(\cosh\theta) \end{aligned}\right)
 \left(\begin{aligned} \sinh\theta P_{l'}^{m'\prime}(\cosh\theta)\hspace{0.5cm} & \frac{m'}{\sinh\theta}P_{l'}^{m'}(\cosh\theta)\\  \frac{m'}{\sinh\theta}P_{l'}^{m'}(\cosh\theta)
 \hspace{0.4cm} & \quad\sinh\theta P_{l'}^{m'\prime}(\cosh\theta) \end{aligned}\right),
\end{split}\end{equation*}which recovers the Casimir interaction energy between a perfectly conducting spherical shell and a perfectly conducting plane \cite{8, 45}.

\section{Small separation asymptotic behavior}

In this section, we consider the asymptotic behavior of the Casimir interaction energy when $d\ll R$, where $d=L-R$ is the distance between the spherical plasma sheet and the planar plasma sheet. Let $$\vep=\frac{d}{R}$$be the dimensionless parameter, and we consider $\vep\ll 1$. There are also another two length parameters in the problem: $1/\Omega_s$ and $1/\Omega_p$. Let
$$\varpi_s=\Omega_s d,\hspace{1cm}\varpi_p=\Omega_pd.$$ They are dimensionless and we assume that they have order 1, i.e.,
$$\varpi_s\sim 1,\hspace{1cm}\varpi_p\sim 1.$$

First we consider the proximity force approximation to the Casimir interaction energy, which approximates the Casimir interaction energy by summing  the local Casimir energy density between two planes over the surfaces.

The Casimir interaction energy density between two planar plasma sheets with respective parameters $\Omega_1$ and $\Omega_2$  is given by the Lifshitz formula \cite{50}:
\begin{equation*}
\mathcal{E}_{\text{Cas}}^{\parallel}(d)=\frac{\hbar c}{4\pi^2}\int_0^{\infty}  d\kappa\int_{0}^{\infty} dk_{\perp} k_{\perp} \left[\ln\left(1-r_{\text{TE}}^{(1)}r_{\text{TE}}^{(2)}e^{-2d\sqrt{k_{\perp}^2+\kappa^2}}\right)+\ln\left(1-r_{\text{TM}}^{(1)}r_{\text{TM}}^{(2)}e^{-2d\sqrt{\kappa^2+k_{\perp}^2}}\right)\right].
\end{equation*}Here $d$ is the distance between the two planar sheets,
\begin{align*}
r_{\text{TE}}^{(i)}=&\frac{\Omega_i}{\Omega_i+\sqrt{\kappa^2+k_{\perp}^2}},\\
r_{\text{TM}}^{(i)}=&-\frac{\Omega_i\sqrt{\kappa^2+k_{\perp}^2}}{\Omega_i\sqrt{\kappa^2+k_{\perp}^2}+\kappa^2}
\end{align*}are nothing but the components of the $\mathbb{T}_2^{\mathbf{k}_{\perp}}$ given in \eqref{eq2_19_4}.

The proximity force approximation for the Casimir interaction energy between a sphere and a plate is then given by
\begin{align*}
E_{\text{Cas}}^{\text{PFA}}=&R^2\int_0^{2\pi} d\phi\int_0^{\pi}d\theta\sin\theta \mathcal{E}^{\parallel}_{\text{Cas}}\left(L+R\cos\theta\right)\\
\sim &2\pi R\int_d^{\infty} du  \mathcal{E}^{\parallel}_{\text{Cas}}(u)\\
=&-\frac{\hbar c R}{2\pi} \int_0^{\infty}  d\kappa\int_0^{\infty} dk_{\perp} k_{\perp}\int_d^{\infty}du\sum_{n=1}^{\infty}\frac{1}{n}\left(
\left[r_{\text{TE}}^{(1)}r_{\text{TE}}^{(2)}\right]^n+\left[r_{\text{TM}}^{(1)}r_{\text{TM}}^{(2)}\right]^n\right)e^{-2un\sqrt{\kappa^2+k_{\perp}^2}}\\
=&-\frac{\hbar c R}{4\pi } \int_0^{\infty}  d\kappa\int_0^{\infty} dk_{\perp} \frac{k_{\perp}}{\sqrt{\kappa^2+k_{\perp}^2}} \sum_{n=1}^{\infty}\frac{1}{n^2}\left(
\left[r_{\text{TE}}^{(1)}r_{\text{TE}}^{(2)}\right]^n+\left[r_{\text{TM}}^{(1)}r_{\text{TM}}^{(2)}\right]^n\right)e^{-2dn\sqrt{\kappa^2+k_{\perp}^2}}\\
=&-\frac{\hbar c R}{4\pi } \int_0^{\infty}  d\kappa\int_0^{\infty} dk_{\perp} \frac{k_{\perp}}{\sqrt{\kappa^2+k_{\perp}^2}} \left(\text{Li}_2\left(r_{\text{TE}}^{(1)}r_{\text{TE}}^{(2)}
e^{-2d\sqrt{\kappa^2+k_{\perp}^2}}\right)+\text{Li}_2\left(r_{\text{TM}}^{(1)}r_{\text{TM}}^{(2)}
e^{-2d\sqrt{\kappa^2+k_{\perp}^2}}\right)\right)\\
=&-\frac{\hbar c R}{4\pi }\int_{0}^{\infty} dq \int_0^{q}  d\kappa \left(\text{Li}_2\left(r_{\text{TE}}^{(1)}r_{\text{TE}}^{(2)}
e^{-2dq}\right)+\text{Li}_2\left(r_{\text{TM}}^{(1)}r_{\text{TM}}^{(2)}
e^{-2dq}\right)\right).
\end{align*}Here $\displaystyle \text{Li}_2(z)=\sum_{n=1}^{\infty} \frac{z^n}{n^2}$ is a polylogarithm function of order 2.  Making a change of variables $dq=t$ and $\kappa=q\sqrt{1-\tau^2}= t\sqrt{1-\tau^2}/d$, we finally obtain
\begin{equation}\label{eq3_25_2}
E_{\text{Cas}}^{\text{PFA}}=-\frac{\hbar c R}{4\pi d^2 }\int_{0}^{\infty} dt \,t \int_0^{1}  \frac{d\tau\,\tau}{\sqrt{1-\tau^2}} \left(\text{Li}_2\left(r_{\text{TE}}^{(1)}r_{\text{TE}}^{(2)}
e^{-2t}\right)+\text{Li}_2\left(r_{\text{TM}}^{(1)}r_{\text{TM}}^{(2)}
e^{-2t}\right)\right),
\end{equation}where
 \begin{align*}
r_{\text{TE}}^{(i)}=&\frac{\Omega_i}{\Omega_i+q}=\frac{\varpi_i}{\varpi_i+t},\\
r_{\text{TM}}^{(i)}=&-\frac{\Omega_iq}{\Omega_iq+ \kappa^2}=-\frac{\varpi_i}{\varpi_i+t(1-\tau^2)}.
\end{align*}

Next, we consider the small separation asymptotic behavior of the Casimir interaction energy up to the next-to-leading order term in $\vep$ from the functional representation  \eqref{eq2_19_2}. In \cite{15}, we have considered the small separation asymptotic expansion of the Casimir interaction between a magnetodielectric sphere and a magnetodielectric plane. Our present scenario is similar to the one considered in \cite{15}. The major differences are the boundary conditions on the sphere and the plate that are encoded in the two matrices $\mathbb{T}_{lm}$ and $\widetilde{\mathbb{T}}_{\mathbf{k}_{\perp}}$. Hence, we do not repeat the calculations that have been presented in \cite{15}, but only present the final result and point out the differences.

The leading term and next-to-leading term of the Casimir interaction energy $E_{\text{Cas}}^0$ and $E_{\text{Cas}}^1$ are given  respectively by
\begin{equation}\label{eq3_13_3}\begin{split}
E_{\text{Cas}}^0= &-\frac{\hbar c R}{4\pi  d^2}\sum_{s=0}^{\infty}\frac{1}{(s+1)^2} \int_0^{\infty}dt\,t \int_0^1 \frac{d\tau\,\tau}{\sqrt{1-\tau^2}} e^{-2t(s+1)}
\sum_{* =\text{TE}, \text{TM}}\left[T^{* }_{0}\widetilde{T}^{*}_{0}\right]^{s+1},
\end{split}\end{equation}\begin{equation}\label{eq3_18_1}\begin{split}
 E_{\text{Cas}}^1  =&-\frac{\hbar c }{4\pi d}\sum_{s=0}^{\infty}\frac{1}{(s+1)^2} \int_0^{\infty}dt\,t \int_0^1 \frac{d\tau\,\tau}{ \sqrt{1-\tau^2}}e^{-2t(s+1)} \Biggl\{\sum_{* =\text{TE}, \text{TM}}\left[T^{* }_{0}\widetilde{T}^{*}_{0}\right]^{s+1}  \left(\mathscr{A}+\mathscr{C}^{*}+\mathscr{D}^{*}\right)+  \mathscr{B}
 \Biggr\}.
\end{split}\end{equation}Here
\begin{align*}
T^{\text{TE} }_{0}=&\frac{ \varpi_s}{ \varpi_s+ t },\\
 T^{\text{TM} }_{0}=& \frac{ \varpi_s}{ \varpi_s+ t\left(1-\tau^2\right)},\\
 \widetilde{T}^{\text{TE} }_{0}=&\frac{ \varpi_p}{ \varpi_p+ t },\\
 \widetilde{T}^{\text{TM} }_{0}=& \frac{ \varpi_p}{ \varpi_p+ t\left(1-\tau^2\right)},
\\
\mathscr{A}=&\frac{ t\tau^2}{3}\left((s+1)^3 +2(s+1)\right)+ \frac{1}{3}\left((\tau^2-2)(s+1)^2-3\tau (s+1)+2\tau^2-1\right)\\
&+\frac{\tau^4+\tau^2-12}{12t\tau^2}(s+1)+\frac{(1+\tau)(1-\tau^2)}{2t\tau^2}-\frac{ (1-\tau^2)}{3t }\frac{1}{s+1},\\
\mathscr{B}=&\frac{ 1-\tau^2}{2t\tau^2  }\left\{\left(T^{\text{TE} }_{0}\widetilde{T}_0^{\text{TM}}+T^{\text{TM} }_{0}\widetilde{T}_0^{\text{TE}}\right)
\frac{\left[T^{\text{TE} }_{0}\widetilde{T}^{\text{TE} }_{0}\right]^{s+1 }
-\left[T^{\text{TM} }_{0}\widetilde{T}^{\text{TM} }_{0}\right]^{s+1} }{T_0^{\text{TE}}\widetilde{T}_0^{\text{TE}}-
T_0^{\text{TM}}\widetilde{T}_0^{\text{TM}}}\right.\\&\left.\hspace{2cm}+2T^{\text{TE} }_{0}\widetilde{T}_0^{\text{TE}}T^{\text{TM} }_{0}\widetilde{T}_0^{\text{TM}} \frac{\left[T^{\text{TE} }_{0}\widetilde{T}^{\text{TE} }_{0}\right]^{s }
-\left[T^{\text{TM} }_{0}\widetilde{T}^{\text{TM} }_{0}\right]^{s} }{T_0^{\text{TE}}\widetilde{T}_0^{\text{TE}}-
T_0^{\text{TM}}\widetilde{T}_0^{\text{TM}}}\right\},\\
\mathscr{C}^{*}=& C_{V }  \mathcal{K}^{*}_1+C_J \mathcal{W}^{*}_1,\\
\mathscr{D}^{*}=&D_{VV}  \mathcal{K}^{*2}_1+D_{VJ}  \mathcal{K}^{* }_1
 \mathcal{W}^{*}_1+D_{JJ} \mathcal{W}^{*2}_1+ D_V \mathcal{K}^{*}_2 +   D_J \mathcal{W}^{*}_2+ (s+1) \mathcal{Y}^{*}_2,
\end{align*}with
\begin{align*}
C_{V}=&-\frac{ \tau}{3}\left((s+1)^3+2(s+1)\right)+\frac{1-\tau^2}{6t\tau}(s+1)^2+\frac{1}{2t}(s+1)+\frac{1-4\tau^2}{12t\tau},\\
C_J=&-\frac{t\tau}{3}\left((s+1)^3-(s+1)\right)+\frac{1}{6\tau}\left((s+1)^2-1\right),\\
D_{VV}=&\frac{1}{12t}\left((s+1)^3-2(s+1)^2+2(s+1)-1\right),\\
D_{JJ}=&\frac{t}{12}\left((s+1)^3-2(s+1)^2-(s+1)+2\right),\\
D_{VJ}=&\frac{1}{6}\left((s+1)^3- (s+1) \right),\\
D_{V}=&\frac{1}{6t}\left(2(s+1)^2 +1\right),\\
D_{J}=&\frac{t}{3}\left( (s+1)^2- 1\right),
\\
\mathcal{K}^{\text{TE}}_{1}=& - \frac{t\tau }{ \varpi_p+ t },\\
\mathcal{K}^{\text{TE}}_{2}=&-\frac{t\left(  \varpi_p+t\left(1-2\tau^2\right)\right)}{2\left( \varpi_p+t \right)^2},\\
\mathcal{K}^{\text{TM}}_{1}=&  \frac{t\left(1-\tau^2\right) }{ \varpi_p+ t\left(1-\tau^2\right)},\\
\mathcal{K}^{\text{TM}}_{2}=& \frac{t\left(1-\tau^2\right)\left( \varpi_p \left(1-2\tau^2\right)+ t\left(1-\tau^2\right) \right)}{2\left( \varpi_p+ t\left(1-\tau^2\right)\right)^2},
\\
\mathcal{W}_{ 1}^{\text{TE}}=&-\frac{\tau}{\varpi_s+ t },\\
\mathcal{W}_{ 2}^{\text{TE}}=&-\frac{\left(t(1-3\tau^2)+\varpi_s\left(1-\tau^2\right)\right)}{2t\left(\varpi_s+t\right)^2},\\
\mathcal{Y}_{ 2}^{\text{TE}}=&-\frac{\tau}{2\left(\varpi_s+t\right)}+\frac{1}{t}\left(\frac{1}{4}-\frac{5\tau^2}{12}\right),\\
\mathcal{W}_{ 1}^{\text{TM}}=&\frac{\tau(1-\tau^2)}{ \varpi_s+t\left(1-\tau^2\right)},\\
\mathcal{W}_{ 2}^{\text{TM}}=&\frac{\left(1-\tau^2\right)\left(t(1-\tau^2)^2+\varpi_s(1-3\tau^2)\right)}{2t\left(\varpi_s+ t\left(1-\tau^2\right)\right)^2},\\
\mathcal{Y}_{ 2}^{\text{TM}}=&\frac{\tau\left(1-\tau^2\right)}{2\left( \varpi_s+t\left(1-\tau^2\right)\right)}+\frac{1}{t}\left(\frac{1}{4}+\frac{7\tau^2}{12}\right).
\end{align*}We have replaced the $l$ in \cite{15} with $t\tau/\vep$. The definition of $\mathscr{B}$, $\mathscr{D}$, $C_V, C_J, D_{VV}, D_{VJ}, D_{JJ}$ are slightly different than those in \cite{15}. For $*$=TE or TM, $\mathcal{K}_1^*, \mathcal{K}_2^*, \mathcal{W}_1^*, \mathcal{W}_2^*$ and $\mathcal{Y}_2^*$ are obtained from the asymptotic expansions of $\mathbb{T}_{lm}$ and $\widetilde{\mathbb{T}}_{\mathbf{k}_{\perp}}$. Hence, there are different than those obtained in \cite{15}.

Using polylogarithm function, we can rewrite the leading term $E_{\text{Cas}}^0$ \eqref{eq3_13_3} as
\begin{equation}\label{eq2_25_1}\begin{split}
E_{\text{Cas}}^0= &-\frac{\hbar c R}{4\pi  d^2} \int_0^{\infty}dt\,t    \int_0^1 \frac{d\tau\,\tau}{\sqrt{1-\tau^2}}
\left(\text{Li}_2\left( T^{\text{TE} }_{0}\widetilde{T}^{\text{TE}}_{0} e^{-2t}\right)+\text{Li}_2\left( T^{\text{TM} }_{0}\widetilde{T}^{\text{TM}}_{0} e^{-2t}\right)\right).
\end{split}\end{equation}It is easy to see that this coincides with the proximity force approximation \eqref{eq3_25_2} when $\varpi_s=\varpi_1$ and $\varpi_p=\varpi_2$.

Notice that the leading term $E_{\text{Cas}}^0$ can be split into a sum of TE and TM contributions. However, because of the $\mathscr{B}$-term, the next-to-leading order term $E_{\text{Cas}}^1$ \eqref{eq3_18_1} cannot be split into TE and TM contributions.

In the limit $\varpi_p, \varpi_s\rightarrow\infty$ which corresponds to perfectly conducting boundary conditions on the sphere and the plate, we find that for $*$=TE or TM,
$\mathcal{K}^{*}_{1}, \mathcal{K}^{*}_{2}, \mathcal{W}^{*}_{1}, \mathcal{W}^{*}_{2}$ vanishes, $T^{*}_{0}=\widetilde{T}^{*}_0=1$,
\begin{align*}
\mathcal{B}=&\frac{ \left(1-\tau^2\right)}{2t\tau^2  }(4s+2),\\
\mathcal{Y}_{ 2}^{\text{TE}}=&\frac{1}{t}\left(\frac{1}{4}-\frac{5\tau^2}{12}\right),\\
\mathcal{Y}_{ 2}^{\text{TM}}=& \frac{1}{t}\left(\frac{1}{4}+\frac{7\tau^2}{12}\right).
\end{align*}
Hence,
\begin{align*}
E_{\text{Cas}}^0= &-\frac{\hbar c R}{2\pi  d^2}\sum_{s=0}^{\infty}\frac{1}{(s+1)^2} \int_0^{\infty}dt\,t \int_0^1 \frac{d\tau\,\tau}{\sqrt{1-\tau^2}} e^{-2t(s+1)}\\
=&-\frac{\hbar c R}{8\pi  d^2}\sum_{s=0}^{\infty}\frac{1}{(s+1)^4} \\
=&-\frac{\hbar c\pi^3 R}{720  d^2},
\\
E_{\text{Cas}}^1  =&-\frac{\hbar c }{4\pi d}\sum_{s=0}^{\infty}\frac{1}{(s+1)^2} \int_0^{\infty}dt\,t \int_0^1 \frac{d\tau\,\tau}{ \sqrt{1-\tau^2}}e^{-2t(s+1)}  \left(2\mathscr{A}+   \mathscr{B}+(s+1)\mathcal{Y}_2^{\text{TE}}+(s+1)\mathcal{Y}_2^{\text{TM}}\right)\\
=&-\frac{\hbar c }{4\pi d}\sum_{s=0}^{\infty}\frac{1}{(s+1)^2} \left(\frac{1}{6(s+1)^2}-\frac{2}{3}\right)\\
=&E_{\text{Cas}}^0\left(\frac{1}{3}-\frac{20}{\pi^2}\right)\frac{d}{R}.
\end{align*}These recover the results for the case where  both the sphere and the  plane are perfectly conducting \cite{35}.

\begin{figure}[h]
\epsfxsize=0.5\linewidth \epsffile{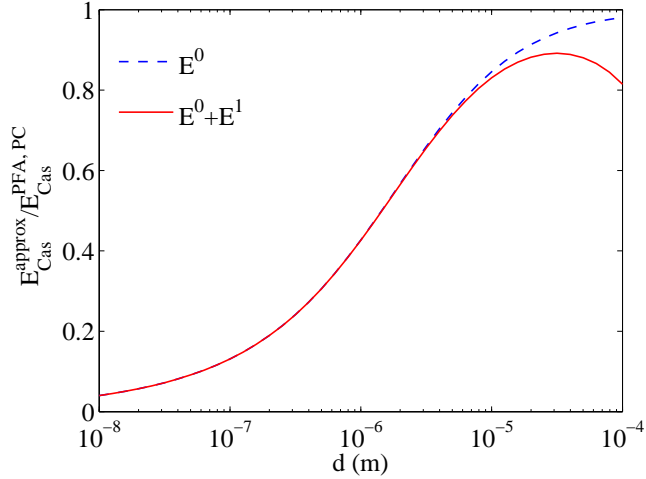} \caption{\label{f1} The   leading order term of the Casimir interaction energy  normalized by $E_{\text{Cas}}^{\text{PFA,PC}}$ (dashed line) and the sum of the leading and next-to-leading order terms normalized by $ E_{\text{Cas}}^{\text{PFA,PC}}$ (solid line) in the case both sphere and plane are graphene sheets.   }\end{figure}

\begin{figure}[h]
\epsfxsize=0.5\linewidth \epsffile{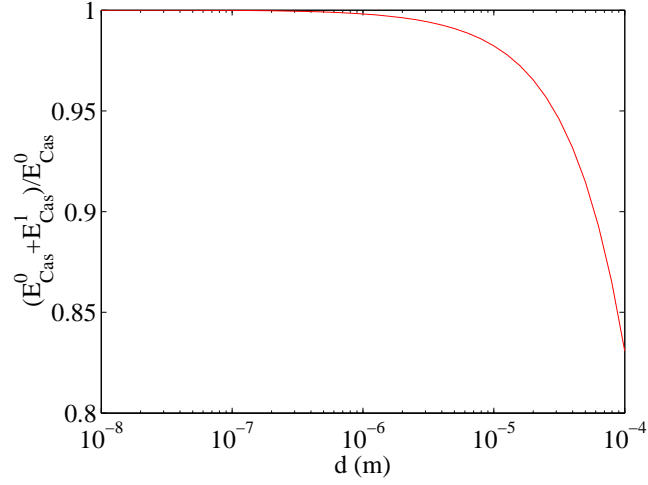} \caption{\label{f2} The  ratio of the sum of the leading and next-to-leading order terms to the leading order term  in the case both sphere and plane are graphene sheets.   }\end{figure}

\begin{figure}[h]
\epsfxsize=0.5\linewidth \epsffile{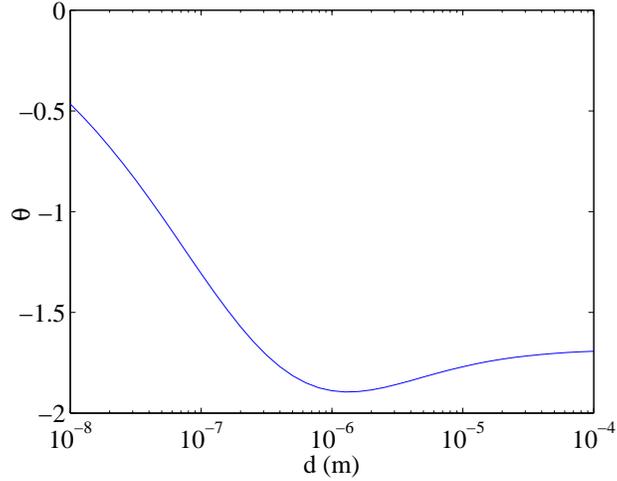} \caption{\label{f3} $\theta$ as a function of $d$ in the case both sphere and plane are graphene sheets.  }\end{figure}

Next, we consider the special case where we have a spherical  graphene sheet in front of a planar graphene sheet. The parameters $\Omega_s$ and $\Omega_p$ are both equal to $6.75\times 10^5 \text{m}^{-1}$ (see Ref. \cite{50}). Assume that the radius of the spherical graphene sheet is $R=1$mm.  Let
\begin{align*}
E_{\text{Cas}}^{\text{PFA,PC}}=-\frac{\hbar c\pi^3 R}{720 d^2}
\end{align*}be the leading term of the Casimir interaction between a perfectly conducting sphere and a perfectly conducting plane. In Fig. \ref{f1}, we plot the ratio of the leading term of the Casimir interaction energy $E_{\text{Cas}}^0$ to $E_{\text{Cas}}^{\text{PFA,PC}}$, and the ratio of the sum of the leading term and next-to-leading order term $\left(E_{\text{Cas}}^0+
E_{\text{Cas}}^1\right)$ to $E_{\text{Cas}}^{\text{PFA,PC}}$. The ratio of $\left(E_{\text{Cas}}^0+
E_{\text{Cas}}^1\right)$  to $E_{\text{Cas}}^0$ is plotted in Fig. \ref{f2}. From these graphs, we can see that the next-to-leading order term plays  a significant correction when $d/R\sim 0.1$.

Another important quantity that characterize the correction to proximity force approximation is
\begin{align*}
\theta=\frac{E^{1}_{\text{Cas}}}{E_{\text{Cas}}^0}\frac{R}{d},
\end{align*} so that
\begin{align*}
E_{\text{Cas}}=E_{\text{Cas}}^0\left(1+\frac{d}{R}\theta+\ldots\right).
\end{align*}In case of perfectly conducting sphere and plane, $\theta$ is a pure number given by \cite{35}:
\begin{align}\label{eq2_27_1}\theta=\frac{1}{3}-\frac{20}{\pi^2}=-1.69.\end{align}
In Fig. \ref{f3}, we plot $\theta$ as a function of $d$ for a spherical graphene sheet in front of a planar graphene sheet. We observe that its variation pattern is significantly different from the case of gold sphere and gold plane modeled by plasma model and Drude model which we studied in \cite{15}. Nevertheless, as $d$ is large enough, $\theta$ approaches the limiting value \eqref{eq2_27_1}.

\begin{figure}[h]
\epsfxsize=0.7\linewidth \epsffile{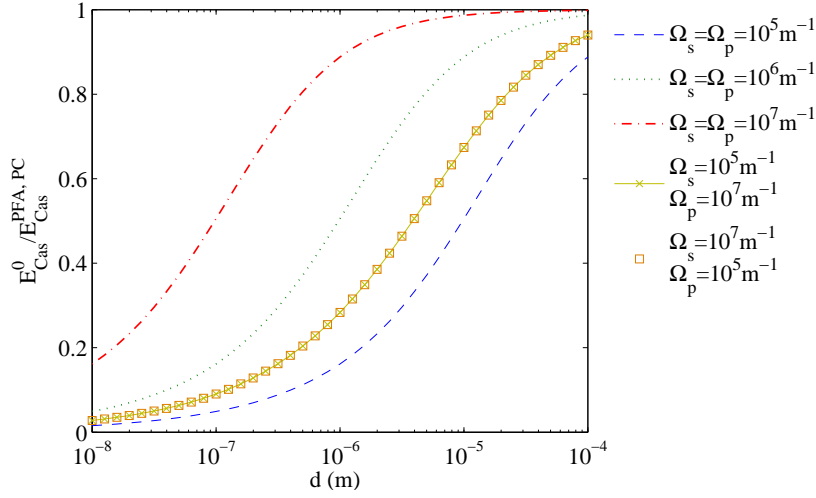} \caption{\label{f4} $E_{\text{Cas}}^0/E_{\text{Cas}}^{\text{PFA,PC}}$ as a function of $d$.  }\end{figure}

\begin{figure}[h]
\epsfxsize=0.7\linewidth \epsffile{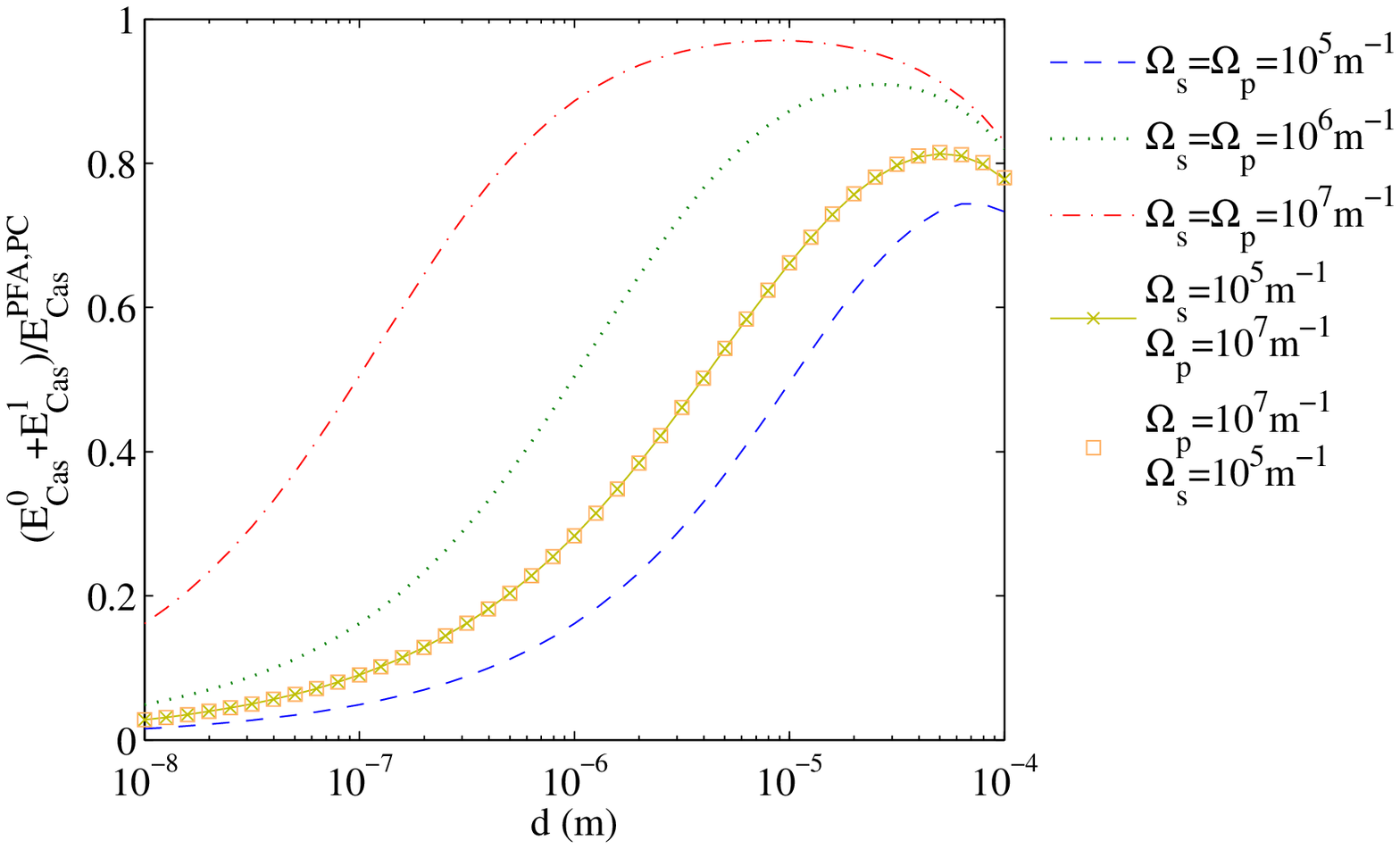} \caption{\label{f5} $(E_{\text{Cas}}^0+E_{\text{Cas}}^1)/E_{\text{Cas}}^{\text{PFA,PC}}$ as a function of $d$.  }\end{figure}

\begin{figure}[h]
\epsfxsize=0.7\linewidth \epsffile{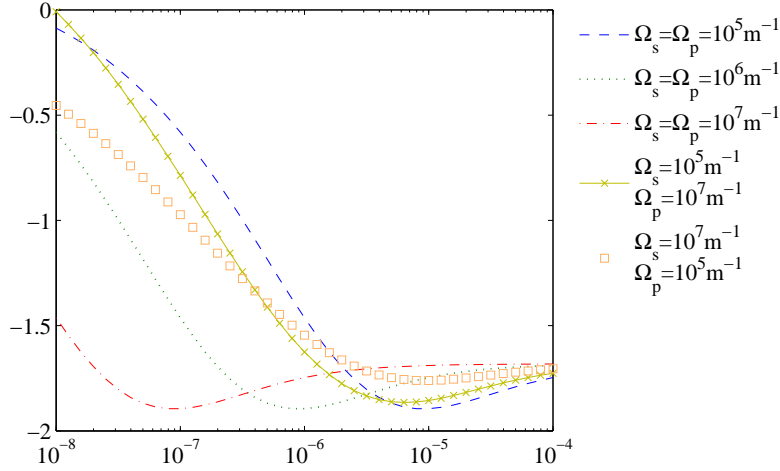} \caption{\label{f6} $\theta$ as a function of $d$.  }\end{figure}

To study the dependence of the Casimir interaction energy on the parameters $\Omega_s$ and $\Omega_p$, we plot in Fig. \ref{f4} and Fig. \ref{f5} respectively the ratio $E_{\text{Cas}}^0/E_{\text{Cas}}^{\text{PFA,PC}}$ and the ratio $(E_{\text{Cas}}^0+E_{\text{Cas}}^1)/E_{\text{Cas}}^{\text{PFA,PC}}$ as a function of $d$ for various values of $\Omega_s$ and $\Omega_p$. The variation of $\theta$ is plotted in Fig. \ref{f6}. It is observe that the larger $\Omega$ is, the larger is the Casimir interaction energy.

The behavior of $\theta$ shown in Fig. \ref{f6} is more interesting. It is observe that it has a minimum which appears at $d\sim\Omega^{-1}$ when $\Omega_s=\Omega_p=\Omega$.

\vfill\pagebreak
\section{Conclusion}

We study the Casimir interaction between a spherical object and a planar object that are made of materials that can be modeled as plasma sheets. The functional representation of the Casimir interaction energy is derived. It is then used to study the small separation asymptotic behavior of the Casimir interaction. The leading term of the Casimir interaction is confirmed to be agreed with the proximity force approximation. The analytic formula for the next-to-leading order term is computed based on a previously established perturbation analysis \cite{15}. The special case where the spherical object and planar object are graphene sheets are considered. The results are found to be quite different from the case of metallic sphere-plane configuration when the separation between the sphere and the plane is small. This may suggest a new experimental setup to test the Casimir effect. It also has potential application to nanotechnology.

\begin{acknowledgments}\noindent
  This work is supported by the Ministry of Higher Education of Malaysia  under   FRGS grant FRGS/1/2013/ST02/UNIM/02/2. I would like to thank M. Bordag for proposing this question.
\end{acknowledgments}

\end{document}